\documentclass[prb,aps,showpacs,twocolumn,10pt]{revtex4-1}

\usepackage{amssymb}

\usepackage{graphicx}
\usepackage[T1]{fontenc}
\usepackage[american]{babel}

\newcommand{\lvosvo}{LaVO$_3$/SrVO$_3$}

\begin{document}

\title{Microstructure and interface studies of \lvosvo\ superlattices}

\author{P.~Boullay}
\author{A.~David}
\author{W.C.~Sheets}
\author{U.~Lüders}
\author{W.~Prellier}
\affiliation{CRISMAT, CNRS UMR 6508, 6 Bd Maréchal Juin, F-14050 Caen Cedex 4, France.} 

\author{H.~Tan}
\author{J.~Verbeeck}
\author{G.~Van~Tendeloo}
\affiliation{EMAT, University of Antwerp, Groenenborgerlaan 171, B-2020 Antwerpen, Belgium.}

\author{C.~Gatel}
\affiliation{CEMES, CNRS UPR 8011, 29 rue Jeanne Marvig, F-31055 Toulouse Cedex 4, France.}

\author{G.~Vincze}
\author{Z.~Radi}
\affiliation{Technoorg Linda Scientific Technical Development Ltd., Rozsa utca 24, 1077 Hungary.}

\author{ }
\affiliation{ }
                  
\date{\today}

\bigskip

\begin{abstract}
\quad The structure and interface characteristics of (LaVO$_3$)$_{6m}$(SrVO$_3$)$_m$ superlattices deposited on (100)-SrTiO$_3$ (STO) substrate were studied using Transmission Electron Microscopy (TEM). Cross-section TEM studies revealed that both LaVO$_3$ (LVO) and SrVO$_3$ (SVO) layers are good single crystal quality and epitaxially grown with respect to the substrate. It is evidenced that LVO layers are made of two orientational variants of a distorted perovskite compatible with bulk LaVO$_3$ while SVO layers suffers from a tetragonal distortion due to the substrate induced stain. Electron Energy Loss Spectroscopy (EELS) investigations indicate changes in the fine structure of the V L$_{23}$ edge, related to a valence change between the LaVO$_3$ and SrVO$_3$ layers.  
\end{abstract}


\maketitle

\section{Introduction}

Oxide superlattices constitute a group of materials, where an artificial periodicity of ultra-thin films made of two different oxides is created. The properties of the superlattices will be governed by the interplay of a number of different effects as the reduced dimensions, the distortion of the lattice due to the mismatch between the constituents and the orbital physics. In such systems, the physical properties of interfaces can be quite surprising as seen for the complex behavior of polar discontinuous interfaces, showing a high mobility two dimensional (2D) electron gas \cite{Oht04,Thi06,Hot07,Bas08,Huijben}, magnetic effects at the interface of non-magnetic materials \cite{Bri07} or even superconductivity \cite{Rey07}. Most of these effects were observed at the SrTiO$_3$/LaAlO$_3$ interface. Another large group of systems showing polar discontinuous interface, i.e. LaBO$_3$/SrBO$_3$ (001) oriented interfaces with B being a transition metal, received much less interest, although the phase diagrams of the solid solutions can indeed be very complex, showing interesting effects as colossal magnetoresistance and charge ordering\cite{Rao98}, metal-to-insulator transitions\cite{Ima98} and complete spin-polarization of the conducting electrons\cite{Par98}. 

\begin{figure}[ht]
\begin{center}
\includegraphics*[width=0.45\textwidth]{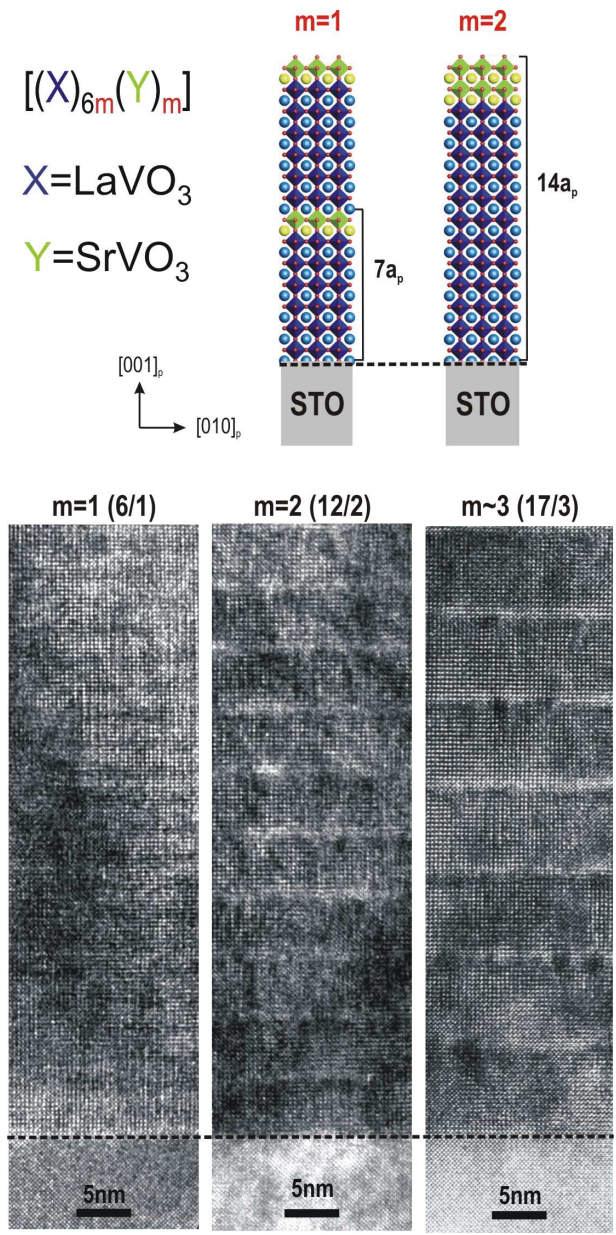}
\caption{\it\small 
Upper: (color online) schematic drawing showing the first two terms of a [(X)$_{6m}$(Y)$_m$] series of superlattice. The composition is identical in both cases but the periodicity of the superlattice is different. 
Lower: HREM images (JEOL 2010F) for the first three terms of the superlattices (LaVO$_3$)$_{6m}$(SrVO$_3$)$_m$ i.e. samples m=1, m=2 and close to m=3 with a compound 17/3 (denoted m$\sim$3).
} 
\label{lvo-svo:MET-HRgen}
\end{center}
\end{figure}

Recently, we concentrated on LaVO$_3$/SrVO$_3$ superlattices since the properties of the parent compounds and of the solid solution show peculiar transitions, which would allow to distinguish between an interface and a 3D effect. For example, LaVO$_3$ is a Mott insulator, showing an antiferromagnetic transition at 143K and a structural transition at 141K \cite{Miy03}. SrVO$_3$ on the other hand is a cubic perovskite, non-magnetic and metallic over all the measured temperature range \cite{Gia95}. The solid solution La$_{1-x}$Sr$_x$VO$_3$ shows a metal-to-insulator transition at x $\approx$ 0.2 \cite{Say75,Ina95, Miy00}. A series of epitaxial (LaVO$_3$)$_{6m}$(SrVO$_3$)$_m$ superlattices having the same nominal composition as La$_{6/7}$Sr$_{1/7}$VO$_3$, a Mott-Hubbard insulator, were grown \cite{Sheets07,Sheets09} and their superlattice period varied (see Fig.\ref{lvo-svo:MET-HRgen}a as an illustration). When m=1, the transport properties vary from bulk-like La$_{6/7}$Sr$_{1/7}$VO$_3$  insulator \cite{Sheets07,Sheets09} to an interesting metallic phase \cite{Luders09} depending on the deposition parameters.
An increase in the periodicity of the superlattice (i.e. increase of the m value) results in metallic samples \cite{Sheets07,Sheets09}.
In this paper, using Transmission Electron Microscopy (TEM), we report the microstructural and interface studies of some of the (LaVO$_3$)$_{6m}$(SrVO$_3$)$_m$ superlattices having the same properties as those reported in the papers by Sheets et al.\cite{Sheets07,Sheets09}.

\section{Experimental}

The samples were prepared by Pulsed Laser Deposition on (100)-SrTiO$_3$ substrates, details are described elsewhere \cite{Sheets07}. X-ray diffraction showed a cube-on-cube epitaxy of the superlattices with a slightly tetragonally distorted perovskite structure (a$_c$/c = 0.388nm/0.395nm = 0.98, where a$_c$ is the in-plane lattice parameter of the pseudo-cubic representation and c the out-of-plane lattice parameter). A large number of orders of superlattice satellites at the main Bragg peaks indicates a high structural quality \cite{Sheets09}. Selected samples were sectioned with a diamond powder coated sawing wheel, encapsulated in a titanium frame (3mm disc) and thinned to 50$\mu$m with SiC abrasive paper and fine grain diamond paste. The mechanical pre-processing was followed by chemical cleaning and low-angle, high-energy ion milling (IV4/H/L Technoorg Ltd). The electron transparent samples were achieved with a dedicated low-energy ion mill (7$^{\circ}$,1000eV, 250eV) of Technoorg Gentle Mill series.
Selected Area electron Diffraction (SAED) and High Resolution Electron Microscopy (HREM) images were obtained using a JEOL 2010F microscope from these cross-section specimens.
High Angle Annular Dark Field (HAADF) imaging and Electron Energy Loss Spectroscopy (EELS) were performed on a JEOL 3000F microscope. The EELS spectrometer was a Gatan GIF2000 1K Phosphor system used at an energy dispersion of 0.5ev/channel and an approximate energy resolution of 1eV. The EELS scans in STEM mode were performed across the layers interfaces with a collection angle of 28.6 mrad and a convergence angle of 10.4 mrad. Geometric Phase Analysis (GPA) \cite{Hytch97,Hytch98} were performed using the DigitalMicrograph GPA plugin on HREM images obtained with a Tecnai F20 microscope equipped with a Cs corrector avoiding delocalization effects.

\section{Results and discussion}

\begin{figure}
\begin{center}
\includegraphics*[width=0.48\textwidth]{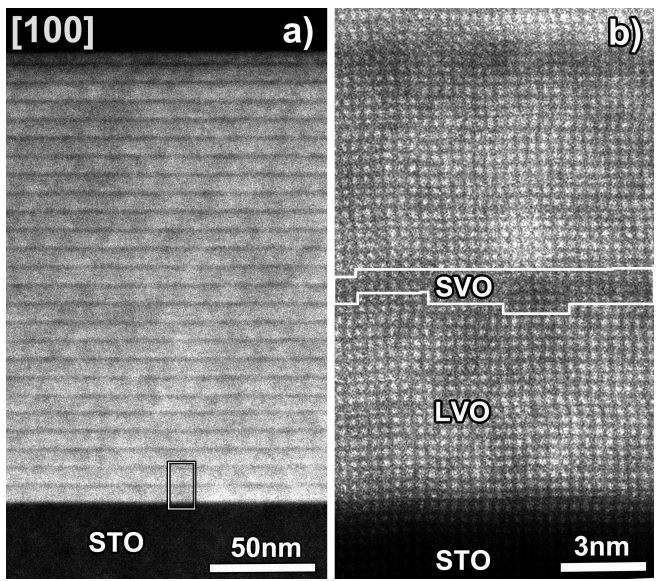}
\caption{\it\small 
a) low magnification HAADF-STEM image of a m=3 (18/3) superlattice. b) High Resolution HAADF-STEM image at the film/substrate interface. The SrVO$_3$ (SVO) layer is marked with its darker contrast compare to the surrounding LaVO$_3$ (LVO) layers. One notices some amount of roughness or atomic steps at the LVO/SVO interfaces.}
\label{lvo-svo:MET-Zcontrast}
\end{center}
\end{figure}

Fig.\ref{lvo-svo:MET-HRgen} shows three HREM images recorded on the terms m = 1, 2 and 3 along the [100]$_{\rm{p}}$ orientation of the substrate. The image of the superlattice for m$\sim$3, shows a series of stacked layers with thick and thin parts regularly alternating. 
Although it is tempting to associate the observed difference in contrast to LaVO$_3$ for the thicker part and to SrVO$_3$ for the lighter regions, origin of contrast in HRTEM is not straightforward and depends on the imaging conditions. 
In the high resolution HAADF STEM image obtained on a m=3 sample and presented in Fig.\ref{lvo-svo:MET-Zcontrast}, it is possible to confirm the realisation of a LaVO$_3$/SrVO$_3$ superlattice because of the presence of strontium in the thinner SrVO$_3$ layer and lanthanum in the thicker LaVO$_3$ layer. For this sample m=3, the interface between the blocks of LaVO$_3$ and SrVO$_3$ shows steps of roughly one perovskite unit preserving the continuity of the SrVO$_3$ layer. In comparison, if one looks at the images obtained on samples m<3 (Fig.\ref{lvo-svo:MET-HRgen}), it becomes difficult to assert the formation of a LaVO$_3$/SrVO$_3$ superlattice with flat interfaces notably for the case where m=1.

\begin{figure}
\begin{center}
\includegraphics*[width=0.42\textwidth]{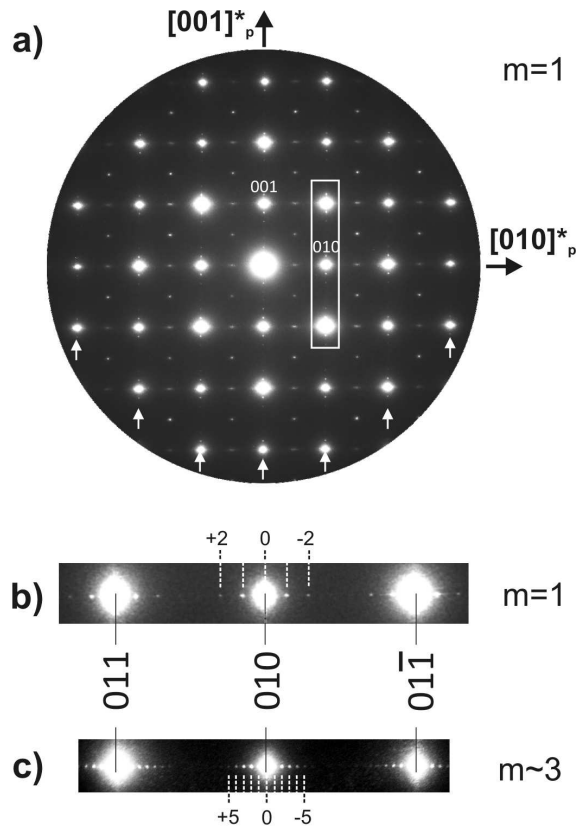}
\caption{\it\small a) [100]$_p$ SAED patterns from a m=1 superlattice. The most intense reflections can be indexed in the perovskite subcell. Along [001]$_p$* (white arrows), satellite reflections appear systematically associated with reflections of the perovksite subcell. b) enlargement of the area delimited by a white rectangular in a). The measured periodicity is 27.1\AA\ $\approx$ 7a$_p$ and is therefore compatible with a m=1 (LaVO$_3$)$_6$/(SrVO$_3$)$_1$ superlattice. c) Enlargement of part of a  [100]$_p$ SAED patterns from the m$\sim$3 (LaVO$_3$)$_{17}$/(SrVO$_3$)$_3$ superlattice with a periodicity of 77.3\AA\ $\approx$ 20a$_p$.
}
\label{lvo-svo:MET-SAEDgen}
\end{center}
\end{figure}

SAED patterns obtained on this m=1 sample (Fig.\ref{lvo-svo:MET-SAEDgen}a) can actually provide more averaged information on the existence of a superlattice than the images presented Fig.\ref{lvo-svo:MET-HRgen}. 
First, the material under investigation (SrTiO$_3$ substrate and LaVO$_3$/SrVO$_3$ film) being structurally related to perovskite, the most intense reflections observed in this diffraction patterns are naturally associated with the perovskite subcell.
Using the [010]$_{\rm{p}}$* direction as reference (a$_{\rm{STO}}$=3.9\AA), the parameter of the perovskite subcell along the [001]$_{\rm{p}}$* direction perpendicular to the substrate is about 3.95\AA\ 
in agreement with the results obtained by X-ray diffraction \cite{Sheets09}.
Along this same [001]$_{\rm{p}}$* direction, one can observe reflections of low intensity and organized on a regular basis on both sides of each of the reflections related to the perovskite subcell (see enlarged area in Fig.\ref{lvo-svo:MET-SAEDgen}b). These reflections can be considered as satellites reflections and indexed using a vector of the form q=(1/$\Lambda$)c*. They are associated with the periodic structure generated by the regular stacking of LaVO$_3$ and SrVO$_3$ in the superlattice. 
It allows us to calculate a super-period $\Lambda$ corresponding to 27.1\AA \ that actually 
corresponds to seven perovskite units as expected for this term m=1. On a second example Fig.\ref{lvo-svo:MET-SAEDgen}c, several orders of sharp satellites spots are observed indicative of the high-quality periodic structure of the deposited superlattices.

\begin{figure}
\begin{center}
\includegraphics*[width=0.40\textwidth]{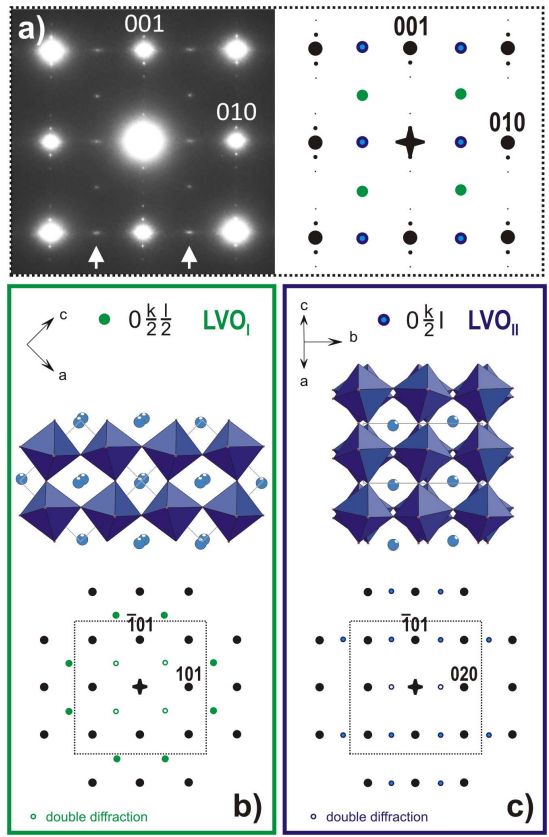}
\caption{\it\small (color online) a) central area of a [100]$_p$ SAED patterns from a m=1 superlattice and its schematic representation. The rows of additional reflections indicated by white arrows can be decomposed into two parts: b) the first located at $0\frac{k}{2}\frac{l}{2}$ positions of the perovskite subcell are compatible with a [010] zone axis patterns for LaVO$_3$ (orthorhombic Pnma bulk reference). c) the second located at $0\frac{k}{2}l$ positions of the perovskite subcell are consistent with a [101] zone axis patterns. }
\label{lvo-svo:MET-SAEDdom}
\end{center}
\end{figure}

\begin{figure}
\begin{center}
\includegraphics*[width=0.46\textwidth]{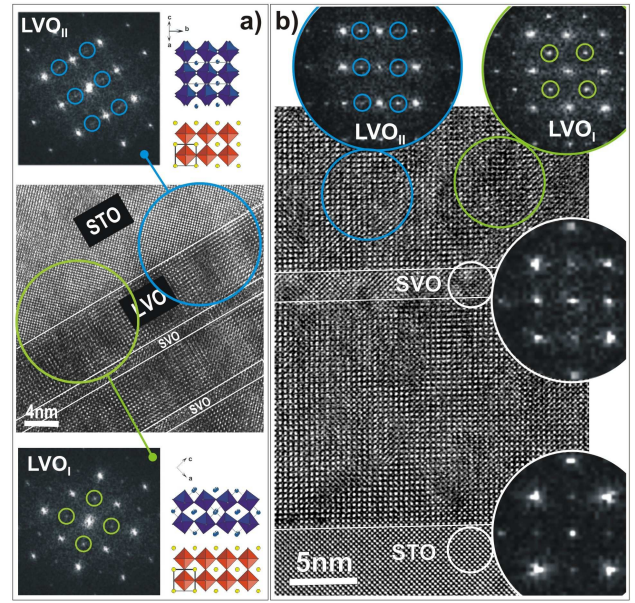}
\caption{\it\small (color online) HREM images of the film/substrate interface for two samples of the series (LaVO$_3$)$_{6m}$(SrVO$_3$)$_m$ with a) m$\sim$3 and b) m=7. For each sample, Fourier transforms performed on the encircled areas show that LaVO$_3$ layers are formed by the imbrication of LVO$_I$ and LVO$_{II}$ orientated domains.}
\label{lvo-svo:MET-HRLVO}
\end{center}
\end{figure}

The diffraction pattern also provides information on the LaVO$_3$ component of the superlattice. When considering the reflections of the perovskite subcell and its satellites (Fig.\ref{lvo-svo:MET-SAEDdom}a), there are indeed extra reflections whose presence can not be explained on the basis of this description. 
The extra rows of reflections, marked by two arrows in Fig.\ref{lvo-svo:MET-SAEDdom}a, can be separated into two subsets located at positions $0\frac{k}{2}l$ and $0\frac{k}{2}\frac{l}{2}$. 
These reflections denote the presence of a superstructure as referred to a simple perovskite lattice. It can be explained if one considers that the LaVO$_3$ component of the superlattice presents a distorted perovskite structure involving rotation of VO$_6$ octahedra consistent with the ones existing in bulk LaVO$_3$ \cite{Bordet93}.
In this case, considering the existence of 90$^\circ$ oriented domains, the reflections $0\frac{k}{2}\frac{l}{2}$ (Fig.\ref{lvo-svo:MET-SAEDdom}b) and $0\frac{k}{2}l$ (Fig.\ref{lvo-svo:MET-SAEDdom}c) may be associated, respectively, to [010] and [101] zone axes of a Pnma structure having cell parameters a$_{\rm{p}}\sqrt{2}\times$2a$_{\rm{p}}\times$a$_{\rm{p}}\sqrt{2}$.
In Fig.\ref{lvo-svo:MET-SAEDdom}, which schematically summarizes the analysis, we see that these two orientations correspond to the case where the b-axis of the Pnma LaVO$_3$ structure
is parallel to the substrate plane but differs by an in-plane rotation of 90$^\circ$. 
Considering its Pnma bulk form, the epitaxial relationship for LVO can be written as (101)LVO$\parallel$(001)STO/SVO with for LVO$_{\rm{I}}$: [010]LVO$\parallel$[100]STO/SVO and for LVO$_{\rm{II}}$: [010]LVO$\parallel$[010]STO/SVO \cite{mono}. 
Notably we do not observe the orientation variant (010)LVO$\parallel$(001)STO/SVO with [101]$\parallel$[010]STO/SVO that would have the b-axis perpendicular to the plane of the substrate (see an illustration in our recent study of Pnma BiCrO$_3$ films deposited on (100)-SrTiO$_3$ \cite{David10}). 
For the various samples observed by TEM and corresponding to values m=1,2,3 and 7, SAED patterns confirm the existence of a perovskite distorted structure with orientation domains for the LaVO$_3$ layer even for m=1 where the LaVO$_3$ layers are about 6 u.c. thick. 
While such a structural features were not reported, to our knowledge, in ultrathin LaVO$_3$ films, we can find similarity with recent works on ultrathin LaMnO$_3$ films \cite{Aruta09} and on LaMnO$_3$/SrMnO$_3$ \cite{Shah10} superlattices, where 10-12 u.c. LaMnO$_3$ layers are shown to adopt a distorted perovskite structure with octahedral tilting comparable to what is observed in bulk LaMnO$_3$ (and so LaVO$_3$).   
In Fig.\ref{lvo-svo:MET-HRLVO}a and b, corresponding to a sample m$\sim$3 and m=7 respectively, HREM images and associated Fourier transforms reveal that the two LVO$_{\rm{I}}$ and LVO$_{\rm{II}}$ orientations are present concurrently in every LaVO$_3$ layer including the first deposited. 
The 90$^\circ$ oriented domains appear on a local scale as indicated by Fourier transform characteristics of the two LVO$_{\rm{I}}$ and LVO$_{\rm{II}}$ variants.
In Fig.\ref{lvo-svo:MET-HRLVO}b, the contrast in the SrVO$_3$ layer appears uniform and the Fourier transform is similar to that of SrTiO$_3$ substrate.

\begin{figure}
\begin{center}
\includegraphics*[width=0.49\textwidth]{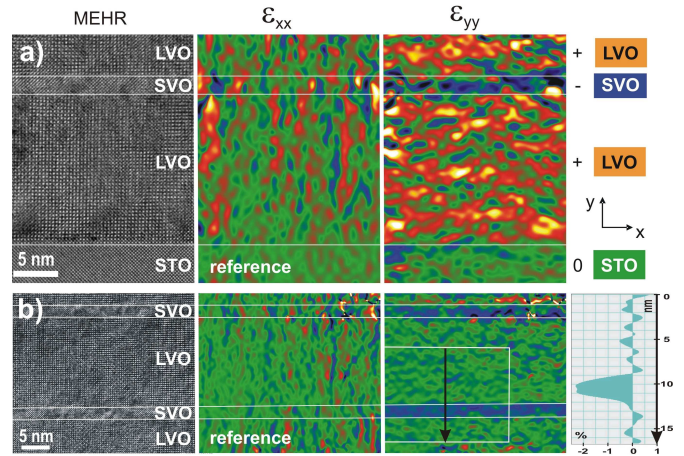}
\caption{\it\small (color online) a) HREM image of the film/substrate interface from a 135nm thick m=7 superlattice and the local deformations obtained by the GPA method along the direction in-plane ($\varepsilon_{xx}$) and out-of-plane ($\varepsilon_{yy}$). Reference is made in the STO substrate. 
Compared to the substrate chosen as a reference area, the green color indicates no change, the blue a negative deformation and the red-orange a positive deformation. 
b) HRTEM image of LVO/SVO interfaces (middle of the film) and the local deformations $\varepsilon_{xx}$ and $\varepsilon_{yy}$ associated. Reference is made this time in the LVO layer. The profile across the $\sim$3nm thick SVO layer indicates a deformation of about -2.2\% .
}
\label{lvo-svo:GPA}
\end{center}
\end{figure}

While it is clear that our superlattices are epitaxially grown onto (100)-SrTiO$_3$, the question to know whether the films are fully strained by the substrate actually remains.
Fig.\ref{lvo-svo:GPA} represents the result of the geometric phase analysis (GPA) performed on a m=7 (LaVO$_3$)$_{6m}$(SrVO$_3$)$_m$ superlattice. In Fig.\ref{lvo-svo:GPA}a, for the deformations along the direction parallel to the substrate ($\varepsilon_{xx}$), we see that there is no difference in contrast between the substrate chosen as reference and the first two LVO/SVO layers. This indicates that both compounds are strained in the plane and they have adopted the parameters of SrTiO$_3$. In Fig.\ref{lvo-svo:GPA}b showing LVO/SVO interfaces, the same phenomenon is observed reflecting the fact that the in-plane strain is kept throughout the film thickness.
If we now look at the images that correspond to out-of-plane deformations ($\varepsilon_{yy}$), we notice a difference. The LaVO$_3$ part is affected by a dilatation in the direction perpendicular to the substrate while SrVO$_3$ is under a compressive strain taking SrTiO$_3$ as a reference.
If one refers to the parameters of the bulk LaVO$_3$ (<a$ _p $>=3.925\AA) and SrVO$_3$ (a=3.843\AA), the relative difference with SrTiO$_3$ (a=3.905\AA) is of -0.5\% and 1.6\%, respectively. We can therefore assume that LaVO$_3$ has a slight compression in the plane which results in an out-of-plane expansion of its lattice spacing, in contrast to the SrVO$_3$ layers where epitaxy requires an in-plane expansion of lattice which induces an out-of-plane compression. GPA is a quantitative method that allows to obtain an estimate of the relative deformation experienced by the SrVO$_3$ layer taking the LaVO$_3$ layer as the reference (Fig.\ref{lvo-svo:GPA}b). The measure indicates that the out-of-plane lattice parameter of the SrVO$_3$ layer is about 2.2\% smaller than the surrounding LaVO$_3$ layers.
It should be noted, however, that we are here within the limit of this type of analysis in the sense that the deformations are relatively small. This explains the noisy appearance of  the deformations images obtained in Fig.\ref{lvo-svo:GPA}. The analysis of smaller terms (m<3) has proven to be even more difficult. Nevertheless, we can estimate that such an achievement for a term m=7 is quite representative of our films.

\begin{figure}
\begin{center}
\includegraphics*[width=0.48\textwidth]{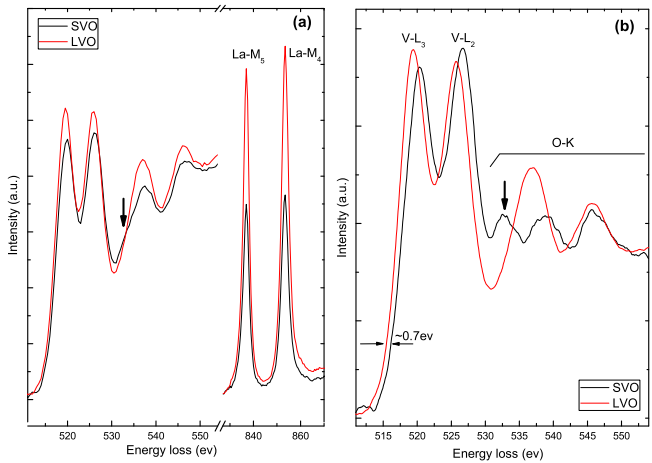}
\caption{\it\small (color online) The probe scanning EELS spectra in STEM mode on SrVO$_3$ and LaVO$_3$ layers from a m=3 (18/3) superlattice. a) SrVO$_3$ and LaVO$_3$ spectra which are background removed before V L$_{23}$ edge and energy aligned by crosscorrelation to the La M$_5$ peak (present in SrVO$_3$ layer as well because of spatial resolution limitations and signal delocalization). b)SrVO$_3$ and LaVO$_3$ spectra (V L$_{23}$ edge and O K edge) after plural scattering removal. The SrVO$_3$ is subtracted by the weighted LaVO$_3$ spectrum to eliminate the effect of overlap. 
}
\label{lvo-svo:EELS}
\end{center}
\end{figure}

Finally, EELS spectra were recorded on a m=3 sample with a STEM probe scanning across the layers. This means that all spectra are collected under the same experimental conditions. Because the SrVO$_3$ layer is very thin and the spatial resolution of the probe is limited ($\approx 1$~nm), the spectrum from the SrVO$_3$ layer shows some residual La signal but the amount is considerably smaller than on the LaVO$_3$ layer (Fig.\ref{lvo-svo:EELS}a). This effect will also cause that the differences in spectra between the two types of layers will be less pronounced. However, the superposition of the LaVO$_3$ signal onto the  SrVO$_3$ spectrum can be partially eliminated by subtracting a weighted LaVO$_3$ spectrum until the La signal is completely removed. Fig.\ref{lvo-svo:EELS}b shows the SrVO$_3$ spectrum after such treatment together with the LaVO$_3$ spectrum (power law background and plural scattering are removed in advance). The V L$_{23}$-edge of SrVO$_3$ is clearly shifted approximately 0.7 eV to higher energies with respect to that of LaVO$_3$. The height ratio of the L$_2$/L$_3$ peak of V is also larger in the SrVO$_3$ spectrum than in the LaVO$_3$ spectrum. Both the chemical shift and the L$_{2,3}$ ratio change demonstrate a higher V valence in the SrVO$_3$ layer as compared to that of the LaVO$_3$ layers\cite{Varela,Dhondt}. Furthermore, comparing to the LaVO$_3$ spectrum (Fig.\ref{lvo-svo:EELS}a), the valley between the V L$_2$ peak and O K edge is shallower for SrVO$_3$ and the first peak of the O K-edge is broader. An extra pre-peak on the O K-edge is clearly separated after removing the LaVO$_3$ signal (marked by a black arrow in Fig.\ref{lvo-svo:EELS}b). The two derived fine structures mainly agree with reference\cite{Kourkoutis}. On the one hand, the first peak of the O K-edge of SrVO$_3$ (Fig.\ref{lvo-svo:EELS}b, indicated by an arrow) shows a further split in reference\cite{Kourkoutis} which is slightly recognizable in our spectrum as well. On the other hand, the width of SrVO$_3$ V L$_{23}$ edge is much broader than that of LaVO$_3$ in reference\cite{Kourkoutis} but they are almost the same for our spectra. Both differences, especially the latter one, may be related to a change in electronic structure at the interface but this is difficult to interpret directly from the spectrum and requires further density functional calculations.

\section{conclusion}

Transmission electron microscopy investigations confirm the general structural quality of the samples and the formation of strained (LaVO$_3$)$_{6m}$(SrVO$_3$)$_m$ superlattices. The layers are homogeneously flat and chemically well defined. The superlattices are found to be epitaxially grown and show satellite reflections in the diffraction pattern in good agreement with the periodicities targeted by varying the m value. At the atomic level however, there are atomic steps at the LaVO$_3$/SrVO$_3$ interfaces which could conduct to discontinuities in the SrVO$_3$ layers for the lower m=2 and m=1 terms. This amount of roughness is however in agreement with the picture of SrVO$_3$ acting as geometrically confined dopant layers\cite{Luders09}.  
For all the investigated superlattices (m=1 to 7), the LaVO$_3$ layers are made of coherently grown 90$^\circ$ oriented domains whose structure present tiltings of the octahedra compatible with what is observed in bulk LaVO$_3$. This assumption is mostly motivated by the similarity between the observed SAED patterns and HREM images with the ones expected for bulk LaVO$_3$. 
Based on GPA results obtained on a m=7 superlattice, the SrVO$_3$ layers appear in compressive strain leading to a tetragonal distortion with a shrinking of the out-of-plane parameter as compared to the bulk. 
EELS investigations indicate changes in the fine structure of the V L$_{23}$ edge, related to a valence change between the LaVO$_3$ and SrVO$_3$ layers. The exact amount of this change  is at present unknown and would require further studies. The existence of a mixed valence in adjacent VO$_2$ sheets with partially localized electrons can not be infert from the present work.
The resolution of the current STEM-EELS experiment is somewhat limited and does not allow to look at the V edge changes within the different unit cells of the SrVO$_3$ layer. 
Moreover, the limited flatness of the interfaces would create a smeared out EELS signature even with a smaller probe size.

\section{acknowledgment}
This work was carried out in the frame of the STREP MACOMUFI (NMP4-CT-2006-313321) supported by the European Commission and by the CNRS, France. Financial support from C'Nano Nord-Ouest (2975) and PHC STAR (21465YL) projects are also acknowledged. P. Boullay acknowledges financial support from the french CNRS and CEA METSA network for the realization of GPA analyses at CEMES. H. Tan would like to acknowledge the financial support from FWO-Vlaanderen (Project nr. G.0147.06). J. Verbeeck also acknowledges financial support from the European Union under the Framework 6 program under a contract for an Integrated Infrastructure Initiative (Reference 026019 ESTEEM).


\begin{thebibliography}{99}
\bibitem{mono} Notice that a strict strained growth of either LVO$_{\rm{I}}$ or LVO$_{\rm{II}}$ on large domains on SrTiO$_3$ would imply a distortion of the orthorhombic cell since the 90$^\circ$ angle between [100] and [001] directions can not be kept considering these epitaxial relations.
\end{thebibliography}
\end{document}